\documentclass[twocolumn,showpacs,preprintnumbers,amsmath,amssymb]{revtex4}

\usepackage{graphicx}
\usepackage{dcolumn}
\usepackage{bm}
\usepackage{epsfig}

\begin{document}


\title{Simple criteria for projective measurements
with linear optics}
\author{Peter van Loock and Norbert L\"{u}tkenhaus}
\affiliation{Quantum Information Theory Group,
Zentrum f\"{u}r Moderne Optik,
Universit\"{a}t Erlangen-N\"{u}rnberg,
91058 Erlangen, Germany}

\begin{abstract}
We derive a set of criteria to decide whether a given
projection measurement can be, in principle, exactly implemented
solely by means of linear optics.
The derivation can be adapted to various detection methods,
including photon counting and homodyne detection.
These criteria enable one to obtain easily No-Go theorems
for the exact distinguishability of orthogonal quantum states with linear
optics including the use of auxiliary photons and conditional
dynamics.
\end{abstract}


\maketitle

\section{Introduction}

Joint orthogonal {\it projection measurements}
are an essential tool in quantum communication. The most prominent
example is the Bell measurement that is used, for instance,
in quantum teleportation \cite{Benn1telep}.
The canonical way to perform these measurements
relies on signal interaction. An example is the optical
interaction of light pulses.
The latter is particularly relevant for practical applications,
since light, traveling
at high speed through an optical fiber and allowing for
an efficient broadband information encoding, is the
most convenient medium for the implementation of
quantum communication protocols.
In discrete-variable implementations based on single photons,
the required strong nonlinear optical interactions
are hard to obtain.
Alternatively, it is a promising
approach to replace interaction by interference,
readily available via {\it linear optics}, and by feedback after
detection. There are important cases, however, where
linear optics is not sufficient to enable specific projective
measurements exactly. For instance, a complete measurement
in the qubit polarization Bell basis is not possible
within the framework of linear optics
including beam splitters, phase shifters, auxiliary photons
and conditional dynamics utilizing photon counting
\cite{NL99,Vaidman99}.
However, using non-trivial entangled states of $n$
auxiliary photons and conditional dynamics, a
perfect projection measurement can be approached
asymptotically with a failure rate scaling as $1/n$ \cite{Knill}
or, in a modified version of the scheme of Ref.~\cite{Knill}
based on similar resources and tools, with an intrinsic error rate
scaling as $1/n^2$ \cite{Franson}.
In any case, No-Go statements for exact implementations always
indicate whenever finite (and cheaper) resources
and less sophisticated tools, such as a fixed
array of linear optics, are not sufficient
for an arbitrarily good efficiency.

In this article, we propose a new approach to the problem
of projective measurements with linear optics and
photon counting. Since orthogonal states remain orthogonal
after linear optical mode transformations, the inability of
exactly discriminating orthogonal states
is due to the measurements in the Fock basis.
In the new approach, we replace the actual detections
by a dephasing of the (linearly transformed) signal states.
In other words, the detection mechanism is mimicked by
destroying the coherence of the signal states and turning
them into mixtures diagonal in the Fock basis.
With the resulting density operators,
the distinguishability is then expressible in terms of
quantum mechanical states.
By considering exact distinguishability, this
yields a hierarchy of simple conditions for a complete
projection measurement.
We give a few examples where we employ these
conditions in order to make general statements
and to derive (some known and some new) No-Go theorems on
linear-optics state discrimination.
Moreover, projection measurements based on
detection schemes other than photon counting
can also be described within the framework of our formalism.
In this respect, we include a brief discussion on
homodyne-detection based quadrature measurements.
However, the essence of our work is
the proposal of a new universal method.
The unified perspective upon which our approach
is based shall open the path to new results and applications
including more general measurements than projective ones.

\section{The criteria}

Let us define the vectors $\vec a=(\hat a_1,\hat a_2,\hdots,\hat a_N)^T$
and $\vec a^\dagger=
(\hat a_1^\dagger,\hat a_2^\dagger,\hdots,\hat a_N^\dagger)$
representing the annihilation and creation operators of all the
electromagnetic modes involved, respectively.
A linear-optics circuit can be described via the
input-output relations $\vec c= U \vec a$ or
$\vec c^\dagger= \vec a^\dagger U^\dagger$ with a
unitary $N\times N$ matrix $U$.
Conversely, the mixing of $N$ optical modes
due to any unitary $N\times N$ matrix is realizable with
beam splitters and phase shifters \cite{Reck}.
This excludes linear mixing between annihilation
and creation operators, as it results from squeezing transformations.
Those require nonlinear optical interactions.
On the Hamiltonian level, arbitrary states $|\chi\rangle$
are unitarily transformed via linear optics such that
\cite{linoptref}
\begin{equation}\label{generallinoptform}
|\chi_H\rangle=\exp(-i\vec a^\dagger H \vec a) |\chi\rangle \;,
\end{equation}
where $H$ is an $N\times N$ Hermitian matrix.

We consider projection measurements that operate on subspaces ${\cal
S}$ of the
Hilbert space defined over some signal modes. The orthogonal projection
measurement is characterized by one-dimensional projectors
$\Pi_k=|s_k\rangle\langle s_k|$ such that
$\langle s_k|s_l\rangle=0$ for $k \neq l$, and the completeness
relation on the subspace ${\cal S}$ is fulfilled as $\sum_k \Pi_k =
\openone_{\cal S}$. In this setting, the problem of
implementing the projection measurement is equivalent to the
unambiguous discrimination of the orthogonal states $|s_k\rangle$.

The state discrimination may be aided by an auxiliary state
$|\psi_{\rm aux}\rangle$ that is supported on auxiliary modes.
The states to be distinguished then are
$\hat\rho_{k,{\rm in}}=
|\chi_k\rangle\langle\chi_k|$ with
$|\chi_k\rangle=|s_k\rangle\otimes|\psi_{\rm aux}\rangle$.
The entire discrimination process now consists of two steps,
$\hat\rho_{\rm in}\rightarrow\hat\rho_H
\rightarrow\hat\rho_H'$,
where the first step is due to linear optics,
$\hat\rho_H\equiv |\chi_H\rangle\langle\chi_H|$.
In the second step, the detection of the output modes
in the Fock basis is mimicked through dephasing,
\begin{eqnarray}\label{dephasingmechanism}
\hat\rho_H\rightarrow \hat\rho_H'=
\frac{1}{(2\pi)^N} \int d\phi^N
e^{-i\vec a^\dagger D \vec a}\hat\rho_H
e^{i\vec a^\dagger D \vec a}\;,
\end{eqnarray}
with $d\phi^N\equiv d\phi_1d\phi_2\hdots d\phi_N$
and the diagonal $N\times N$ matrix $D$,
$(D)_{ij}=\delta_{ij}\phi_i$.
The distinguishability can then be analyzed
on the level of the density operators $\hat\rho_H'$.
Since exact discrimination is considered,
this leads to a huge simplification, as we shall explain now.

In order to decide on the exact distinguishability
of any two states $|\chi_k\rangle=|s_k\rangle\otimes
|\psi_{\rm aux}\rangle$ and $|\chi_l\rangle=|s_l\rangle\otimes
|\psi_{\rm aux}\rangle$, we may
use ${\rm Tr}(\hat\rho_{k,H}'\hat\rho_{l,H}')$,
where $\hat\rho_{k,H}'$ and $\hat\rho_{l,H}'$ are
the corresponding states after linear optics and dephasing.
We obtain the condition for exact distinguishability,
\begin{eqnarray}\label{derivation1}
{\rm Tr}(\hat\rho_{k,H}'\hat\rho_{l,H}')
&=&\frac{1}{(2\pi)^{2N}}\int d\phi^N d\tilde\phi^N
\nonumber\\
&&\times|\langle\chi_k|e^{i\vec a^\dagger H \vec a}
e^{i\vec a^\dagger (D-\tilde D) \vec a}
e^{-i\vec a^\dagger H \vec a}|\chi_l\rangle|^2\nonumber\\
&\stackrel{!}{=}&0\;,
\end{eqnarray}
where $d\tilde\phi^N\equiv d\tilde\phi_1d\tilde\phi_2
\hdots d\tilde\phi_N$ and
$(\tilde D)_{ij}=\delta_{ij}\tilde\phi_i$.
Due to the positivity of the integrand, this is equivalent to
\begin{eqnarray}\label{derivation2}
&&\langle\chi_k|e^{i\vec a^\dagger H \vec a}
e^{i\vec a^\dagger (D-\tilde D) \vec a}
e^{-i\vec a^\dagger H \vec a}|\chi_l\rangle
\nonumber\\
&=&\langle\chi_k|
e^{i\vec c^\dagger (D-\tilde D) \vec c}
|\chi_l\rangle
= 0\,,
\quad\forall \phi_j, \tilde\phi_j\;,
\end{eqnarray}
where the effect of linear optics is now put into
the operators $\vec c = e^{i\vec a^\dagger H \vec a}
\vec a\, e^{-i\vec a^\dagger H \vec a}$ or
$\vec c = U \vec a$. Let us define  $y_j\equiv \phi_j
- \tilde\phi_j$, $j=1...N$.
Since the derivatives of $\langle\chi_k|
e^{i\vec c^\dagger (D-\tilde D) \vec c}
|\chi_l\rangle$ with respect to any relative phases
$y_j,y_{j'},y_{j''},...$
must also vanish, in particular,
at $\vec y = (y_1,y_2,...,y_N)= \vec 0$, we obtain
the set of conditions
for exact state discrimination,
\begin{eqnarray}\label{hierarchy}
\langle\chi_k|\hat c^\dagger_j\hat c_j
|\chi_l\rangle &=& 0\,,
\quad\forall j\;,\\
\langle\chi_k|\hat c^\dagger_j\hat c_j\,
\hat c^\dagger_{j'}\hat c_{j'}
|\chi_l\rangle &=& 0\,,
\quad\forall j,j'\;, \nonumber\\
\langle\chi_k|\hat c^\dagger_j\hat c_j\,
\hat c^\dagger_{j'}\hat c_{j'}\,
\hat c^\dagger_{j''}\hat c_{j''}\cdots
|\chi_l\rangle &=& 0\,,
\quad\forall j,j',j''\;, \nonumber \\
\quad\quad\quad\vdots\quad\quad\quad
&=&\quad\quad\quad \vdots\quad\quad \forall k\neq l\,.
\nonumber
\end{eqnarray}
These conditions are
{\it necessary} for a complete projection
measurement onto the basis $\{|\chi_k\rangle\}$.
However, if the entire set of conditions
is satisfied, this is in general also a {\it sufficient}
condition, since
$e^{i\vec c^\dagger (D-\tilde D) \vec c}$ is an analytic
function of the relative phases $\vec y$.
Note that orthogonality
$\langle \chi_k|\chi_l\rangle=0$, $\forall k \neq l$,
is the ``0th-order condition''.

By exploiting that
$(\hat c^\dagger\hat c)^n$ is of the form
$\sum_{m=1}^n d_m
(\hat c^\dagger)^m \hat c^m$ with some
coefficients $d_m$ and
that $[\hat c^\dagger_j,\hat c_{j'}]=0$
for $j\neq j'$,
the higher-order conditions can be rewritten
in an equivalent normally ordered form, provided the
lower-order conditions are satisfied.
This leads to the  hierarchy of conditions,
\begin{eqnarray}\label{pickj}
\langle\chi_k|\hat c^\dagger_j\hat c_j
|\chi_l\rangle &=& 0\,,
\quad\forall j\;,\\
\langle\chi_k|\hat c^\dagger_j\hat c^\dagger_{j'}
\hat c_j\hat c_{j'}
|\chi_l\rangle &=& 0\,,
\quad\forall j,j'\;, \nonumber\\
\langle\chi_k|\hat c^\dagger_j\hat c^\dagger_{j'}
\hat c^\dagger_{j''}\hat c_j\hat c_{j'}
\hat c_{j''}\cdots
|\chi_l\rangle &=& 0\,,
\quad\forall j,j',j''\;, \nonumber \\
\quad\quad\quad\vdots\quad\quad\quad
&=&\quad\quad\quad \vdots \quad\quad\forall k\neq l\,.
\nonumber
\end{eqnarray}
In this form, one can directly see that
the hierarchy breaks off for higher-order terms
if the number of
photons in the states $\{|\chi_k\rangle\}$ is bounded.
Hence, for finite photon number,
we end up having a finite hierarchy of
necessary and sufficient conditions for
complete projective measurements.
The states of an orthogonal set
$\{|\chi_k\rangle\}$ are, in principle, exactly distinguishable
via a {\it fixed array of linear optics} represented
by $\vec c = U \vec a$,
if and only if these conditions hold
for the complete set of modes.

The subset of conditions referring only to a particular mode operator
$\hat c_j$ represents {\it necessary} conditions for
exact discrimination based on {\it conditional
dynamics} after detecting that mode $j$. They are given by
\begin{equation}
\label{condcond}
\langle\chi_k| \left( \hat c^\dagger_j \right) ^n \left( \hat c_j
\right )^n
|\chi_l\rangle = 0 \;,
\quad\quad\quad\forall n \geq1, \quad \forall k \neq l\,.
\end{equation}
Already the failure to find some $\hat c_j$ fulfilling
Eq.~(\ref{condcond}) means that as soon as one output mode is
selected and measured, this will make exact
discrimination of the states
impossible. Conversely, one may also use the conditions
of Eq.~(\ref{condcond}) in a constructive way.
The recipe is to find one $\hat c_j$ that satisfies
Eq.~(\ref{condcond}), to calculate the corresponding conditional
states of the remaining modes, and to test them for
their distinguishability.
It is instructive to view this in terms of the
partially dephased states. After
dephasing only one mode $j$, we obtain
\begin{eqnarray}\label{dephasedstates2}
\hat\rho_{k,H}'^{(j)}=\sum_m
\frac{P^{(j)}(m|k)}{m!}(\hat a^\dagger_j)^m
|0\rangle_j\,
|c_{k,m}^{(j)}\rangle
\langle c_{k,m}^{(j)}|\,{}_j\langle 0|\,
\hat a_j^m,\nonumber\\
\end{eqnarray}
where $P^{(j)}(m|k)$ is the probability to find $m$ photons in the measured
mode $j$ for given input state $|\chi_k\rangle$,
and $|c_{k,m}^{(j)}\rangle$ is the corresponding
(normalized) conditional state of the remaining modes. Failure to fulfill
Eq.~(\ref{condcond}) implies that the conditional states
$|c_{k,m}^{(j)}\rangle$ form a
non-orthogonal set in $k$ for each fixed combination
of $(m,j)$. For such
sets, we know that a further exact
discrimination is impossible.
We will show now that the condition
in Eq.~(\ref{condcond}) for $n=1$
suffices to reproduce easily
all known No-Go theorems for projective
measurements with linear optics
including auxiliary photons and conditional dynamics.

\section{Examples}

In this section, we present a few examples
that illustrate the simplicity and usefulness
of the criteria derived in the preceding section.
These examples include general statements
on the effect of extra resources to the exact
distinguishability of arbitrary quantum states and
``back of the envelope'' derivations
of some known and some new No-Go theorems.
Among them, the simplest and most
remarkable example will be
that for a pair of orthogonal two-photon states,
because the previously known No-Go results apply
to sets of at least four orthogonal states
(e.g., the Bell states).

We start by investigating the use of auxiliary photons
\cite{CarolloJModOpt}.
Splitting the input modes
into a set of signal and a set of
auxiliary modes allows us to decompose the mode operator $\hat c_j =
\sum_i U_{ji} \hat a_i$ from Eq.~(\ref{condcond})
into two corresponding parts
as (we drop the index $j$)
$\hat c = b_{\rm s}\hat c_{\rm s}+
b_{\rm aux}\hat c_{\rm aux}$,
with real coefficients $b_{\rm s}$ and $b_{\rm aux}$,
so that $\hat c_{\rm s}|{\bf 0}\rangle\otimes |\psi_{\rm aux}\rangle=
\hat c_{\rm aux}|s_k\rangle\otimes |{\bf 0}\rangle=0$.
Now we find
\begin{eqnarray}\label{auxphotons}
\lefteqn{\langle\chi_k|\hat c^{\dagger}\hat c|\chi_l\rangle}
\nonumber\\
&=&b_{\rm s}^2\langle s_k|\hat c_{\rm s}^{\dagger}
\hat c_{\rm s}|s_l\rangle +
b_{\rm s}b_{\rm aux} \langle s_k|
\hat c_{\rm s}|s_l\rangle\langle\psi_{\rm aux}|\hat c_{\rm aux}^{\dagger}|
\psi_{\rm aux}\rangle\nonumber\\
&&+b_{\rm s} b_{\rm aux} \langle s_k|\hat
 c_{\rm s}^{\dagger}|s_l\rangle\langle\psi_{\rm aux}|\hat c_{\rm aux}|
\psi_{\rm aux}\rangle\nonumber\\
&&+b_{\rm aux}^2\langle s_k|s_l\rangle\langle\psi_{\rm aux}|
\hat c_{\rm aux}^{\dagger}
\hat c_{\rm aux}|\psi_{\rm aux}\rangle.
\end{eqnarray}
The last term always vanishes for $k \neq l$, since the $|s_k\rangle$ are
orthogonal. In any situation where
either the signal states or the auxiliary state have a
fixed photon number, the two middle terms vanish,
and the first-order condition
$\langle\chi_k|\hat c^{\dagger}\hat c|\chi_l\rangle =0$
only depends on the signal states alone,
$b_{\rm s}^2\langle s_k|\hat c_{\rm s}^{\dagger}
\hat c_{\rm s}|s_l\rangle =0$, $\forall k \neq l$.
The trivial case $b_{\rm s}=0$
can be omitted without loss of generality.
It is straightforward to extend this derivation
to any order in Eq.~(\ref{condcond})
by inserting a mode operator decomposed into a signal
and an auxiliary part.
{\it Hence for signal states with a fixed photon number,
auxiliary systems never help, and for signal states
with an unfixed number, adding an auxiliary state may
help, but only provided the auxiliary state
has unfixed number too.}

The No-Go theorem for the
qubit Bell states \cite{NL99,Vaidman99},
\begin{eqnarray}\label{Bellstates}
|\Psi_\pm\rangle&=&\frac{1}{\sqrt{2}}(\hat a_1^\dagger
\hat a_4^\dagger \pm
\hat a_2^\dagger \hat a_3^\dagger)\,|{\bf 0}\rangle\,,\nonumber\\
|\Phi_\pm\rangle&=&\frac{1}{\sqrt{2}}(\hat a_1^\dagger
\hat a_3^\dagger \pm
\hat a_2^\dagger \hat a_4^\dagger)\,|{\bf 0}\rangle \,,
\end{eqnarray}
is obtainable now in a very simple way.
In order to check for the existence of a mode $j$
satisfying Eq.~(\ref{condcond}) for $n=1$, let us
again drop the index $j$ and use the ansatz
$\hat c_{\rm s}\propto\nu_1\hat a_1 + \nu_2\hat a_2 +
\nu_3\hat a_3 + \nu_4\hat a_4$,
by defining $U_{ji}\equiv \nu_i$.
We have six conditions
$\langle\chi_k|\hat c^\dagger\hat c
|\chi_l\rangle = 0$ for the pairs $(k,l)$,
\begin{eqnarray}\label{Bellconditions1}
(\Psi_+,\Psi_-),(\Phi_+,\Phi_-):\;
|\nu_1|^2-|\nu_2|^2\mp |\nu_3|^2\pm |\nu_4|^2=0\,,
\nonumber\\
(\Psi_+,\Phi_+),(\Psi_+,\Phi_-):\;
\nu_1\nu_2^*+\nu_3\nu_4^*\pm\nu_1^*\nu_2
\pm\nu_3^*\nu_4=0\,,
\nonumber\\
(\Psi_-,\Phi_+),(\Psi_-,\Phi_-):\;
\nu_3\nu_4^*-\nu_1\nu_2^*\pm\nu_1^*\nu_2
\mp\nu_3^*\nu_4=0\,.
\nonumber\\
\end{eqnarray}
These conditions imply
\begin{eqnarray}\label{Bellconditions2}
(\Psi_+,\Psi_-),(\Phi_+,\Phi_-)\quad
&\Rightarrow&\quad
|\nu_1|^2=|\nu_2|^2,\, |\nu_3|^2=|\nu_4|^2,
\nonumber\\
(\Psi_+,\Phi_+),(\Psi_+,\Phi_-)\quad
&\Rightarrow&\quad
\nu_1\nu_2^*=-\nu_3\nu_4^*,
\nonumber\\
(\Psi_-,\Phi_+),(\Psi_-,\Phi_-)\quad
&\Rightarrow&\quad
\nu_1\nu_2^*=\nu_3\nu_4^*.
\end{eqnarray}
It can be easily seen that these conditions
have only trivial solutions $\nu_i=0$, $\forall i$,
which proves the No-Go theorem for the Bell states
including auxiliary photons and conditional dynamics.

\begin{figure}[tb]
\epsfxsize=1.0in \epsfbox[220 280 400 570]{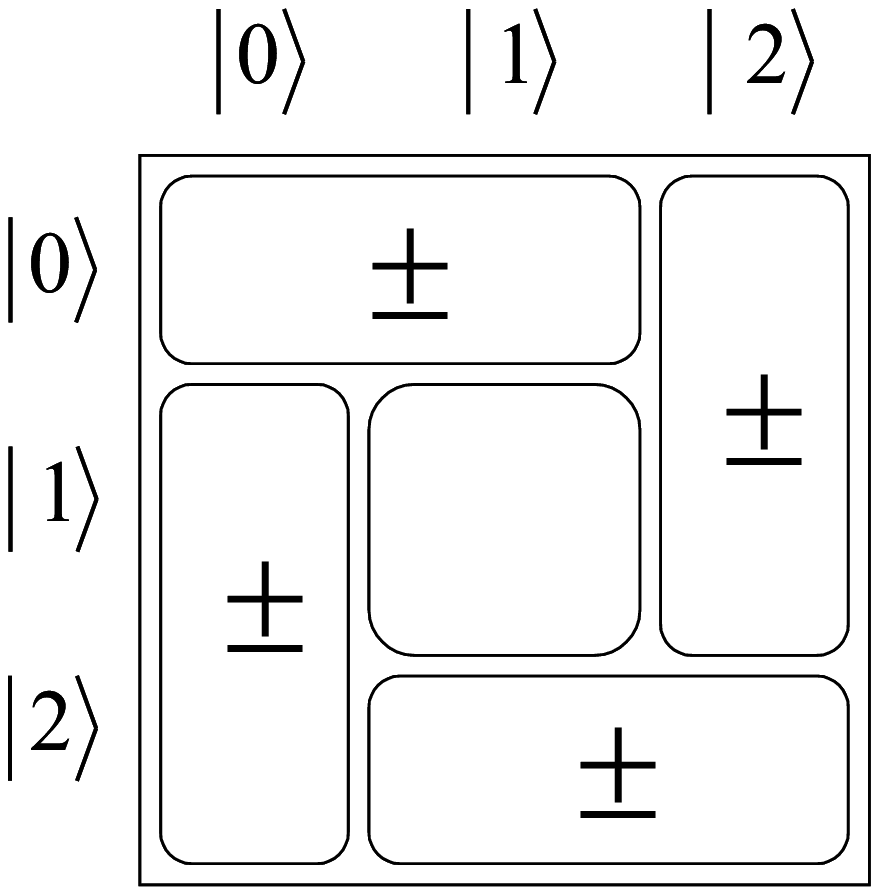}
\caption{\label{fig1}The nine two-qutrit product states which are
undistinguishable via linear optics when encoded into two-photon
states. The three logical basis states $\{|0\rangle,|1\rangle,
|2\rangle \}$ at each side are then represented by a single photon
in one of three modes, for instance, the photonic states
$|s_{1,2}\rangle$ from Eq.~(\ref{qutritstates}) correspond to the
logical states $|0\rangle\otimes (|0\rangle\pm
|1\rangle)/\sqrt{2}$.}
\end{figure}

A similar No-Go theorem
is known \cite{Carollo01} for an orthogonal
set of separable two-qutrit states \cite{Benn99}
\begin{eqnarray}\label{qutritstates}
|s_{1,2}\rangle=\frac{1}{\sqrt{2}} \hat a_1^\dagger
(\hat a_4^\dagger \pm \hat a_5^\dagger)\,|{\bf 0}\rangle,\;
|s_{3,4}\rangle=\frac{1}{\sqrt{2}} \hat a_3^\dagger
(\hat a_5^\dagger \pm \hat a_6^\dagger )\,|{\bf 0}\rangle,
\nonumber\\
|s_{5,6}\rangle=\frac{1}{\sqrt{2}} \hat a_4^\dagger
(\hat a_2^\dagger \pm \hat a_3^\dagger)\,|{\bf 0}\rangle,\;
|s_{7,8}\rangle=\frac{1}{\sqrt{2}} \hat a_6^\dagger
(\hat a_1^\dagger \pm \hat a_2^\dagger )\,|{\bf 0}\rangle,
\nonumber\\
|s_9\rangle = \hat a_2^\dagger \hat a_5^\dagger
\,|{\bf 0}\rangle.\quad\quad\quad\quad\quad\quad\quad\quad\quad
\end{eqnarray}
The entire set of 36 first-order conditions
for one mode $j$ with $\hat c = \sum_i\nu_i \hat a_i $
now leads to
\begin{eqnarray}\label{qutritconditions}
|\nu_1|^2=|\nu_2|^2=|\nu_3|^2,\quad
|\nu_4|^2=|\nu_5|^2=|\nu_6|^2,
\nonumber\\
\nu_1\nu_2^*=\nu_1\nu_3^*=\nu_2\nu_3^*=
\nu_4\nu_5^*=\nu_4\nu_6^*=\nu_5\nu_6^*=0 \; .
\end{eqnarray}
Again, only trivial solutions exist.
Going beyond Ref.~\cite{Carollo01},
we can now easily investigate subclasses
of the set. The full No-Go theorem
also applies to the eight states when
leaving out state $|s_9\rangle$.
For other subclasses, this example illustrates
the role of conditional
dynamics. For instance, leaving out state
$|s_8\rangle$, the conditions remain exactly
those in Eq.~(\ref{qutritconditions}) except that
$|\nu_1|^2$ does not occur in the first line.
The only nontrivial solution is now
where $\nu_1=1$ and $\nu_i=0$, $\forall i=2...6$.
The interpretation is that in order to enable
discrimination of the conditional states
for the entire subset, mode 1 must be
detected first.
This can be seen intuitively in
Eq.~(\ref{qutritstates}) and in Fig.~\ref{fig1}.

With the help of the hierarchy of conditions,
one can now easily find new No-Go theorems.
Consider the orthogonal set of four two-qubit states,
\begin{eqnarray}\label{nonmaxentstates}
|s_1\rangle&=&(\alpha \hat a_1^\dagger \hat a_4^\dagger +
\beta \hat a_2^\dagger \hat a_3^\dagger)\,|{\bf 0}\rangle\,,
\nonumber\\
|s_2\rangle&=&(\beta^* \hat a_1^\dagger \hat a_4^\dagger -
\alpha^* \hat a_2^\dagger \hat a_3^\dagger)\,|{\bf 0}\rangle\,,
\nonumber\\
|s_3\rangle&=&(\gamma \hat a_1^\dagger \hat a_3^\dagger +
\delta \hat a_2^\dagger \hat a_4^\dagger)\,|{\bf 0}\rangle \,,
\nonumber\\
|s_4\rangle&=&(\delta^* \hat a_1^\dagger \hat a_3^\dagger -
\gamma^* \hat a_2^\dagger \hat a_4^\dagger)\,|{\bf 0}\rangle \,.
\end{eqnarray}
If all four states are entangled, $|\alpha\beta|>0$ and
$|\gamma\delta|>0$, only trivial
solutions exist for the six first-order
conditions Eq.~(\ref{condcond}) with $n=1$.
Hence, the full No-Go statement applies including
auxiliary photons and conditional dynamics.
For only two entangled states, e.g.
$|\alpha\beta|>0$ and $\gamma=0$, one mode $\hat c_j$
always exists which satisfies Eq.~(\ref{condcond}).
However, there are only trivial solutions to the
second-order condition in Eq.~(\ref{pickj}) for some
pairs of modes $\hat c_j$ and $\hat c_{j'}$ ($j\neq j'$),
if the two states are {\it nonmaximally} entangled.
In fact, a fixed array of linear optics is not sufficient
in this case, but a conditional-dynamics solution exists.
If the two states are {\it maximally} entangled,
any order in Eq.~(\ref{pickj}) is satisfied with
a $50:50$ beam splitter.

A particularly simple example is the following {\it pair} of
orthogonal states,
\begin{eqnarray}\label{cuteexample}
\frac{1}{\sqrt{2}} (|20\rangle\pm |11\rangle)\;,
\end{eqnarray}
described in the Fock basis.
We find that the
$n=1$ and $n=2$ conditions of Eq.~(\ref{condcond}) can
be simultaneously satisfied only trivially, $\nu_1=\nu_2=0$.
Thus, there is no linear optical discrimination scheme
for the two states of Eq.~(\ref{cuteexample}),
not even with the help of conditional dynamics
and auxiliary photons,
since the two states have fixed photon number.
In fact, this No-Go statement applies to the whole
family of pairs of orthogonal states,
$\alpha |20\rangle + \beta |11\rangle$ and
$\beta^* |20\rangle - \alpha^* |11\rangle$,
for $|\alpha\beta|>0$.

What about quantitative statements beyond the No-Go theorems
for exact state discrimination?
A linear-optics network with photon counting  yields for each input
state a classical probability distribution for the pattern of
photon detections in the output modes. This distribution can be used to
estimate the input state. A possible measure in the context of
estimating an input state is the probability of minimum error
\cite{Fuchsthesis}.
For four equally probable output distributions,
it can be written as
\begin{eqnarray}\label{minerror}
P_{\rm error}^{\rm min}=1 - \frac{1}{4}\sum_i\;
{\rm max}_k[P(i|k)]\;,
\end{eqnarray}
where $P(i|k)$ is the conditional probability for obtaining
the result $i$ (pattern of the photon
detections) given the distribution $k$.
Using the classical distributions of the results $i$
in the totally dephased states with the two-photon Bell
states of Eq.~(\ref{Bellstates}) as the input states
(parametrized by an arbitrary unitary $4\times 4$
matrix $U$), we found numerically that
$P_{\rm error}^{\rm min}\geq 1/4$.
This bound can be attained
by using a $50:50$ beam splitter \cite{Weinfurter},
where
\begin{eqnarray}\label{dephasedstatesBS}
\hat\rho_{\Psi_+,{\rm BS}}'&=&\frac{1}{2}\,
(|1100\rangle\langle 1100|+|0011\rangle\langle 0011|)
\;,\nonumber\\
\hat\rho_{\Psi_-,{\rm BS}}'&=&\frac{1}{2}\,
(|1001\rangle\langle 1001|+|0110\rangle\langle 0110|)
\;,\nonumber\\
\hat\rho_{\Phi_\pm,{\rm BS}}'&=&\frac{1}{4}\,
(|2000\rangle\langle 2000|+|0200\rangle\langle 0200|
\nonumber\\
&&\quad\;+|0020\rangle\langle 0020|+|0002\rangle\langle 0002|)
\;,
\end{eqnarray}
corresponding to the optimal partial Bell measurement
without auxiliary photons and conditional dynamics
\cite{Calsam01}.

\section{Quadrature measurements}

So far, the dephasing approach has been solely used to describe
the decohering effect of photon detections, i.e.,
measurements in the Fock basis.
However, it is worthwhile pointing out that this method is
applicable to other kinds of measurements too.
We may also consider, for example, homodyne detections, i.e.,
measurements in a continuous-variable basis.
In that case, the appropriate replacement
in the dephasing formula of Eq.~(\ref{dephasingmechanism})
is
\begin{eqnarray}\label{dephasingmechanismcv}
e^{i\vec a^\dagger D \vec a}=
e^{i\sum_j\phi_j\hat a^\dagger_j\hat a_j}
\rightarrow e^{i\sum_j\phi_j\hat x_j^{(\theta_j)}}\;,
\end{eqnarray}
where $\hat x_j^{(\theta_j)}=(\hat a_j e^{-i\theta_j}+
\hat{a}_j^{\dagger} e^{+i\theta_j})/2$ are the quadratures
of mode $j$. For example, for $\theta_j=0$ and $\theta_j=\pi/2$,
we obtain respectively the position $\hat x$
and momentum $\hat p$
associated with the mode's harmonic oscillator.
The derivation of a set of necessary and sufficient
conditions for exact state discrimination,
Eqs.~(\ref{derivation1}-\ref{hierarchy}),
also follows through with the replacement in
Eq.~(\ref{dephasingmechanismcv}). The resulting
conditions in that case become (we drop the superscript $\theta_j$)
\begin{eqnarray}\label{hierarchycv}
\langle\chi_k|\hat x_j^c
|\chi_l\rangle &=& 0\,,
\quad\forall j\;,\\
\langle\chi_k|\hat x_j^c\,\hat x_{j'}^c
|\chi_l\rangle &=& 0\,,
\quad\forall j,j'\;,\nonumber\\
\langle\chi_k|\hat x_j^c\,\hat x_{j'}^c\,
\hat x_{j''}^c\cdots
|\chi_l\rangle &=& 0\,,
\quad\forall j,j',j''\;,\nonumber\\
\quad\quad\quad\vdots\quad\quad\quad
&=&\quad\quad\quad \vdots \quad \quad \forall k \neq l\,,
\nonumber
\end{eqnarray}
where
$\hat x_j^c=(\hat c_j e^{-i\theta_j}+
\hat{c}_j^{\dagger} e^{+i\theta_j})/2$
denotes the quadratures of mode $j$
after the linear-optics circuit with
$\vec c = U \vec a$.
A continuous-variable Bell measurement
discriminates between
the two-mode eigenstates of
the relative position $\hat x_1 - \hat x_2$
and total momentum $\hat p_1 + \hat p_2$.
This can be achieved with a simple $50:50$ beam
splitter and subsequent $\hat x$ and $\hat p$
measurements at the two output ports \cite{SamKimble}.
Conditional dynamics is not needed. However,
in order to satisfy the above conditions for all
(that is two) modes,
two conjugate quadratures must be detected, for example,
$\hat x_1^c=(\hat c_1 +
\hat{c}_1^{\dagger})/2=(\hat x_1 - \hat x_2)/\sqrt{2}$
and
$\hat x_2^c=(\hat c_2 -
\hat{c}_2^{\dagger})/2i=(\hat p_1 + \hat p_2)/\sqrt{2}$.
Here, $\hat x_j$ and $\hat p_j$ are the two conjugate
quadratures of the input modes $\hat a_j$.
Hence, due to the orthogonality of the continuous-variable
Bell states, the described scheme represents a solution to
the above conditions.
In a very intuitive way, this explains why a fixed
linear-optics scheme suffices to perform a continuous-variable
Bell measurement with arbitrarily high efficiency, in contrast
to a qubit Bell measurement:
the continuous-variable Bell states are eigenstates of the
detected quadratures, whereas the qubit Bell states
are no eigenstates of the detected photon numbers.

\section{Summary and outlook}

In summary, we have presented a new approach to describe
the processing of quantum states via linear optics
including photon counting or
other measurements such as homodyne detection.
The advantage of this approach is that the detection
mechanism is included in the transformation from the input
quantum states to the output quantum states.
For the case of a complete projection measurement
onto a (joint) orthogonal basis,
we obtained a hierarchy of necessary and sufficient conditions.
When photon counting is considered,
this hierarchy breaks off and yields a finite set of
simple conditions for states with finite photon numbers.
Apart from homodyne detection,
our universal approach
can also be used to include other
``continuous-variable tools'' such as
displacments and squeezing.
It also provides a promising method to treat
more general scenarios, e.g. the realization of
general measurements (POVM's) with linear optics
\cite{JohnPRA02}.
Any POVM can be described via Naimark extension
as an orthogonal von Neumann measurement
in a larger Hilbert space.
The extended signal states may then be analyzed
using the criteria derived in this paper.
This generalization is particularly significant,
because it would extend our approach from qualitative
statements on exact projection measurements
to quantitative statements on approximate
projection measurements.

Although progress is being made in enhancing the effective strength of
nonlinear optical interactions, it appears reasonable to exploit the
entire toolbox of linear optics first and explore it, in order to be aware
of its capabilities, but also its limitations.
In the recent work of Ref.~\cite{Knill}, the authors
demonstrate that the capabilities of linear optics are unexpectedly
broad, however, unfeasibly many extra resources may be needed for a
good performance. We hope that the question of the trade-off
between these extra resources and the performance
can be attacked utilizing our criteria.

\acknowledgments

We are grateful to John Calsamiglia and Bill Munro for
useful comments. We also acknowledge
the financial support of the DFG under
the Emmy-Noether programme, the EU FET network
RAMBOQ (IST-2002-6.2.1), and the network of
competence QIP of the state of Bavaria (A8).

\end{document}